\documentstyle[emulateapj,onecolfloat,psfig]{article}

\newcommand{\da}{d_A}

\newcommand{\tot}{{\rm t}}
\newcommand{\hal}{{\rm h}}
\newcommand{\Poi}{{\rm P}}
\newcommand{\lin}{{}}

\newlength{\tskip}
\newlength{\colwidth}

\newcommand{\beq}{\begin{equation}}
\newcommand{\eeq}{\end{equation}}
\newcommand{\beqa}{\begin{eqnarray}}
\newcommand{\eeqa}{\end{eqnarray}}

\def\simgt{\gtrsim}
 \newcommand{\wj}{\left(
                          \begin{array}{ccc}
                          l_1  &  l_2  & l_3 \\
                            0  &  0    &  0
                          \end{array}
                          \right)}
\newcommand{\wjm}{\left(
                          \begin{array}{ccc}
                          l_1  &  l_2  & l_3 \\
                           m_1  &  m_2   &  m_3
                          \end{array}
                          \right)}
\newcommand{\bi}{B_{l_1 l_2 l_3}}

\newcommand{\rad}{r}    

\begin{document}
\twocolumn[
\title{Weak Gravitational Lensing Bispectrum}
\author{Asantha Cooray$^1$ and Wayne Hu$^{2,}$\altaffilmark{3}}
\affil{
$^1$Department of Astronomy and Astrophysics, University of Chicago,
Chicago, IL 60637\\
$^2$Institute for Advanced Study, Princeton, NJ 08540\\
E-mail: asante@hyde.uchicago.edu, whu@ias.edu}
\submitted{Submitted for publication in The Astrophysical Journal}


\begin{abstract}
Weak gravitational lensing observations probe the spectrum
and evolution of density fluctuations and the cosmological parameters
which govern them. The non-linear evolution of large scale
structure produces a non-Gaussian signal which is potentially
observable in galaxy shear data. We study the three-point statistics of
the convergence, specifically the bispectrum, 
using the dark matter halo approach which 
describes the density field in terms of correlations between and
within dark matter halos.
Our approach allows us to study the effect of the mass distribution in observed
fields, in particular the bias induced by the lack of rare massive halos
(clusters) in observed fields.   We show the convergence skewness
is primarily due to rare and massive dark matter halos with skewness
converging to its mean value only if halos of mass $M > 10^{15} M_{\sun}$ are 
present.  This calculational method can in principle be used to correct for
such a bias as well as to search for more robust statistics related
to the two and three point correlations. 
\end{abstract}

]

\altaffiltext{3}{Alfred P. Sloan Fellow}

\section{Introduction}

Weak gravitational lensing of faint galaxies probes the
distribution of matter along the line of sight.  Lensing by
large-scale structure (LSS) induces
correlation in the galaxy ellipticities at the percent level
(e.g., \cite{Blaetal91} 1991; \cite{Mir91} 1991; 
\cite{Kai92} 1992).  Though challenging to measure, these
correlations
provide important cosmological information that is
complementary to that supplied by
the cosmic microwave background and potentially as precise
(e.g., \cite{JaiSel97} 1997;
\cite{Beretal97} 1997; \cite{Kai98} 1998; \cite{Schetal98}
1998; \cite{HuTeg99} 1999; \cite{Coo99} 1999; \cite{Vanetal99} 1999;
see \cite{BarSch00} 2000 for a recent review).
Indeed several recent studies have provided the first clear evidence
for weak lensing in so-called blank fields (e.g., \cite{Vanetal00}
2000; \cite{Bacetal00} 2000; \cite{Witetal00} 2000; \cite{Kaietal00} 2000),
though more work is
clearly needed to understand even the statistical errors 
(e.g. \cite{Cooetal00b} 2000b).
 
Given that weak gravitational lensing results from the projected mass
distribution, the statistical properties of weak lensing convergence
reflect those of the dark matter. 
Non-linearities in the mass distribution induce non-Gaussianity
in the convergence distribution.
With the growing observational and theoretical 
interest in weak gravitational lensing, 
statistics such as the skewness 
have been suggested as probes of 
cosmological parameters and 
the non-linear evolution of large scale structure
(e.g., \cite{Beretal97} 1997; 
\cite{JaiSelWhi00} 2000; \cite{Hui99} 1999; \cite{MunJai99} 1999; \cite{Vanetal99}
1999). 

Here, we extend previous studies by considering the full convergence bispectrum, 
the Fourier space analog of three-point function. The bispectrum contains all the
information present at the three point level, whereas conventional
statistics, such as skewness, do not.  The calculation of the 
convergence bispectrum requires detailed
knowledge of the dark matter density bispectrum, which can be
obtain analytically through perturbation theory (e.g.,
\cite{Beretal97} 1997)
or numerically through simulations (e.g., \cite{JaiSelWhi00} 2000; \cite{WhiHu99}
1999). Perturbation theory, however, is not applicable at all scales of
interest, while numerical simulations are limited by computational expense
to a handful of realizations of cosmological models with modest dynamical
range.  
Here, we use a new approach to obtain the density
field bispectrum analytically by describing the underlying three point
correlations as due to contributions from (and correlations between)
individual dark matter halos.

Techniques for studying the dark matter density field through
halo contributions have recently been developed 
(\cite{Sel00} 2000; \cite{MaFry00b} 2000b; \cite{Scoetal00} 2000) 
and applied to two-point lensing statistics
(\cite{Cooetal00b} 2000b).
The critical ingredients are: the Press-Schechter formalism (PS;
\cite{PreSch74} 1974) for the mass function; the NFW
profile of \cite{Navetal96} (1996), and the halo bias
model of \cite{Moetal97} (1997). 
The dark matter
halo approach provides a physically motivated method to calculate the
bispectrum.   By calibrating the halo parameters with N-body simulations,
it can be made accurate across the scales of interest.
Since lensing probes scales
ranging from linear to deeply non-linear, this is an important advantage
over perturbation-theory calculations.

Throughout this paper, we will take $\Lambda$CDM as our fiducial cosmology 
with parameters $\Omega_c=0.30$ for the CDM density,
$\Omega_b=0.05$ for the baryon density, $\Omega_\Lambda=0.65$ for the
cosmological constant, $h=0.65$ for the dimensionless Hubble
constant and a scale invariant spectrum of
primordial fluctuations, normalized to  galaxy  cluster abundances 
($\sigma_8=0.9$ see \cite{ViaLid99} 1999)  
and consistent with COBE (\cite{BunWhi97} 1997).
For the linear power spectrum, we take the
fitting  formula for the transfer function given in \cite{EisHu99} (1999).

In \S \ref{sec:density}, we review the dark matter
halo approach to modeling the density field.  In \S \ref{sec:convergence}
we apply the formalism to the convergence power spectrum, skewness, and
bispectrum.  We summarize our results in \S \ref{sec:conclusions}.

\section{Density Power Spectrum and Bispectrum}
\label{sec:density}
\subsection{General Definitions}

Underlying the halo approach is the assertion that dark matter halos of virial mass $M$ are 
locally biased tracers of density perturbations in the linear regime.  In this
case, functional relationship between the overdensity of halos and mass can be expanded 
in a Taylor series
\begin{equation}
\delta^\hal({\bf x},M;z) =
b_1(M;z)\delta({\bf x};z)   + {1 \over 2} b_2(M;z)\delta^2({\bf x};z)  + \ldots 
\end{equation}
\cite{Moetal97} (1997) give the following analytic predictions for the bias parameters
which agree well with simulations:
\begin{equation}
b_1(M;z) = 1 + \frac{\nu^2(M;z) - 1}{\delta_c} \, ,
\end{equation}
and 
\begin{eqnarray}
b_2(M;z) &=& \frac{8}{21}[b_1(M;z)-1] + { \nu^2(M;z) -3 \over \sigma^2(M;z)}\,.
\end{eqnarray}
Here $\nu(M,z) = \delta_c/\sigma(M,z)$, where
$\sigma(M,z)$ is the rms fluctuation within a top-hat filter at the
virial radius corresponding to mass $M$,
and $\delta_c$ is the threshold overdensity of spherical
collapse (see \cite{Hen00} 2000 for useful fitting functions).

Roughly speaking, the perturbative aspect of the 
clustering of the dark matter is described by the correlations between 
halos,
whereas the nonlinear aspect is described by the correlations within halos,
i.e. the halo profiles.
We will consider the Fourier analogies of the 2 and 3 point correlations of
the density field defined in the usual way
\begin{equation}
\left< \delta^*({\bf k}) \delta({\bf k}') \right> = (2\pi)^3
\delta({\bf k}-{\bf k}') P^\tot(k) \, ,
\end{equation}
\begin{equation}
\nonumber
\left< \delta({\bf k}_1) \delta({\bf k}_2) \delta({\bf k}_3) \right> =
(2\pi)^3 \delta({\bf k}_1+{\bf k}_2 + {\bf k}_3) B^\tot(k_1,k_2,k_3)
\, .
\end{equation}
Here and throughout, we occasionally suppress the redshift dependence where no 
confusion will arise.

As we shall see, 
these spectra are related to the {\it linear} density power spectrum $P(k)$ through 
the bias parameters and the normalized 3d Fourier transform of the halo density profile $\rho(r,M)$
\begin{equation}
y(k,M) = \frac{1}{M} \int_0^{r_v} dr\, 4 \pi r^2 \rho(r,M) \frac{\sin (k
r)}{kr} \, . 
\end{equation}
Note that $y(k,M)$ can be written as a combination of 
sine and cosine integrals for computational purposes and
$y(k,M) \rightarrow 0$ as $k \rightarrow 0$.

It is convenient then to define a general integral over the halo mass function
$dn/dM$,
\begin{eqnarray}
I_\mu^\beta(k_1,\ldots,k_\mu;z) &\equiv&
\int dM \left(\frac{M}{\rho_b}\right)^\mu \frac{dn}{dM}(M,z) b_\beta(M)  \nonumber\\
&& \times y(k_1,M)\ldots y(k_\mu,M)\,, 
\end{eqnarray}
where $b_0 \equiv 1$.
We use the Press-Schechter (PS; \cite{PreSch74} 1974) 
mass function to describe $dn/dM$. We take the minimum mass
to be $10^3$ M$_{\sun}$ while the maximum mass is varied to
study the effect of massive halos on lensing convergence statistics.
In general, masses above $10^{16}$ M$_{\sun}$ do not contribute to low
order statistics due to the exponential decrease in the number 
density of such massive halos.

\begin{figure*}[t]
\centerline{\psfig{file=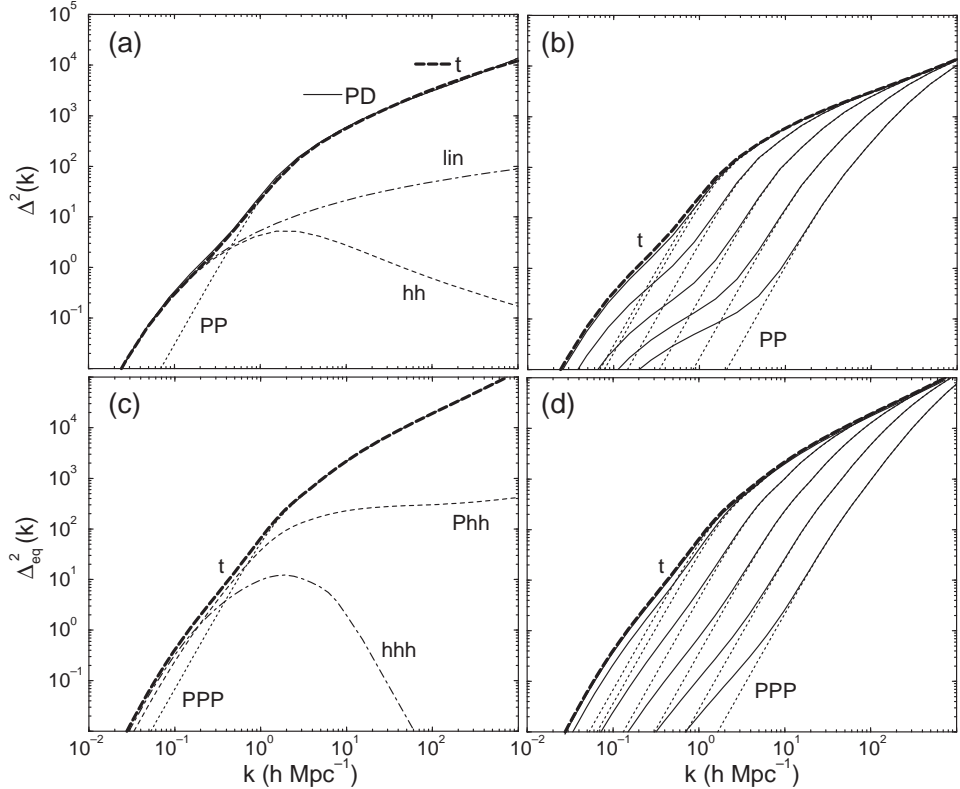,width=5in}}
\caption{Dark matter density spectra under the halo prescription. 
(a) Dark matter power spectrum at the present. 
(b) Mass cut off effects on the power spectrum. 
(c) Equilateral bispectrum at the present.
(d) Mass cut off effects on the equilateral bispectrum. 
The power spectrum is compared with the PD fitting function and the
linear $P(k)$ in (a).
In (b) and (d), from bottom to top,
the maximum mass used in the calculation is $10^{11}$,
$10^{12}$, $10^{13}$, $10^{14}$, $10^{15}$ and $10^{16}$ M$_{\sun}$.}
\label{fig:dmpower}
\end{figure*}
\subsection{Power Spectrum and Bispectrum}
Following \cite{Sel00} (2000), we can decompose the density 
power spectrum, as a function of redshift, into contributions
from single halos (shot noise or ``Poisson'' contributions), 
\begin{equation}
P^{\Poi\Poi}(k) = I_2^0(k,k) \,,
\end{equation}
 and correlations between two halos, 
\begin{equation}
P^{\hal\hal}(k) = \left[  I_1^1(k) \right]^2 P^\lin(k)\,,
\end{equation}
such that 
\begin{equation}
P^\tot = P^{\Poi\Poi} +  P^{\hal\hal} \,.
\end{equation}
As $k \rightarrow 0$, $P^{\hal\hal} \rightarrow P(k)$.

Similarly, we decompose the bispectrum into
terms involving one, two and three halos 
(see \cite{SchBer91} 1991; \cite{MaFry00b} 2000b):
\begin{eqnarray}
B^\tot &=& B^{\Poi\Poi\Poi}  + B^{\Poi\hal\hal}+  B^{\hal\hal\hal} \, ,
\end{eqnarray}
where 
\begin{eqnarray}
B^{\Poi\Poi\Poi}(k_1,k_2,k_3) = I_3^0(k_1,k_2,k_3)\,,
\end{eqnarray}
for single halo contributions,
\begin{eqnarray}
B^{\Poi\hal\hal}(k_1,k_2,k_3) = I_2^1(k_1,k_2) I_1^0(k_3) P^\lin (k_3) + 
{\rm Perm.}
\label{eqn:BPhh}
\end{eqnarray}
for double halo contributions, and 
\begin{eqnarray}
B^{\hal\hal\hal}(k_1,k_2,k_3) &=& 
\left[ 2 J(k_1,k_2,k_3) I_1^1(k_3) + I_1^2(k_3) \right]
\label{eqn:Bhhh}
\\
&&\times 
I_1^1(k_1) I_1^1(k_2) P^\lin(k_1)P^\lin(k_2) 
+ {\rm Perm.}
\nonumber
\end{eqnarray}
for triple halo contributions.
Here the 2 permutations are $k_3 \leftrightarrow k_1$, $k_2$.  
Second order perturbation theory tells us that
(\cite{Fry84} 1984; \cite{Bouetal92} 1992;
\cite{KamBuc99} 1999)
\begin{eqnarray}
J(k_1,k_2,k_3) &=& 1 - \frac{2}{7}\Omega_m^{-2/63} 
                     + \left( { k_3^2 - k_1^2 - k_2^2 \over 2 k_1 k_2} \right)^2
\nonumber\\
&&\times  \left[ \frac{k_1^2+k_2^2}{k_3^2 - k_1^2 -k_2^2} 
+  \frac{2}{7}\Omega_m^{-2/63}\right] \, .
\end{eqnarray}
As $k \rightarrow 0$, $B^{\hal\hal\hal} \rightarrow B^{\rm PT}$, where
$B^{\rm PT}$ is the bispectrum predicted by second-order perturbation
theory
\begin{eqnarray}
B^{\rm PT}(k_1,k_2,k_3) &=& 
2 J(k_1,k_2,k_3) P^\lin(k_1)P^\lin(k_2) + {\rm Perm.} \, , \nonumber \\
\end{eqnarray}
with permutations following $k_3 \leftrightarrow k_1$, $k_2$.  

\subsection{Halo Profiles}

To apply the halo prescription for the power spectrum and bispectrum,
we need to know the profile of the halos.  We take the NFW profile
(\cite{Navetal96} 1996) with a density distribution 
\begin{equation}
\rho(r,M) = \frac{\rho_s}{(r/r_s)(1+r/r_s)^{2}} \, .
\end{equation}
The density profile can be integrated and related to the total dark
matter mass of the halo within $r_v$
\begin{equation}
M =  4 \pi \rho_s r_s^3 \left[ \log(1+c) - \frac{c}{1+c}\right] \, ,
\label{eqn:mass}
\end{equation}
where the concentration, $c$, is defined as $r_v/r_s$.
Choosing $r_v$ as the virial radius of the halo, spherical
collapse tells us that 
$M = 4 \pi r_v^3 \Delta(z) \rho_b/3$, where $\Delta(z)$ is
the overdensity of collapse (see
e.g. \cite{Hen00} 2000) and $\rho_b$ is the background matter density
today. We use
comoving coordinates throughout.
By equating these two expressions, one can
eliminate $\rho_s$ and describe the halo by its mass $M$ and 
concentration $c$. 

Following \cite{Cooetal00b} (2000b), we take the concentration
of dark matter halos to be
\begin{equation}
c(M,z) = a(z)\left[ \frac{M}{M_*(z)}\right]^{-b(z)}\,,
\label{eqn:concentration}
\end{equation}
where $a(z) = 10.3(1+z)^{-0.3}$ and $b(z) = 0.24(1+z)^{-0.3}$.
Here $M_*(z)$ is the non-linear mass scale at which the peak-height
threshold, $\nu(M,z)=1$.
The above concentration is chosen so that  dark matter halos
provide a reasonable match to the
the non-linear density power spectrum as predicted by the \cite{PeaDod96} (1996);
it extends the treatment of \cite{Sel00} (2000) to the redshifts of interest
for lensing.   We caution the reader that eqn.~(\ref{eqn:concentration}) is only
a good fit for the $\Lambda$CDM model assumed.

\subsection{Results}

In Fig.~\ref{fig:dmpower}(a-b), we show the density field power spectrum
today ($z=0$), written 
such that $\Delta^2(k)=k^3 P(k)/2\pi^2$ 
is the power per logarithmic interval in
wavenumber.
In Fig~\ref{fig:dmpower}(a), we show individual contributions
from the single and double halo terms 
and a comparison to the non-linear power
spectrum as predicted by the \cite{PeaDod96} (1996) fitting function.
In Fig.~\ref{fig:dmpower}(b), we show the dependence of density field power as a
function of maximum mass used in the calculation. 

Since the bispectrum generally scales as the square of the power spectrum,
it is useful to define 
\begin{equation}
\Delta_{\rm eq}^2(k) \equiv \frac{k^3}{2\pi^2} \sqrt{B(k,k,k)} \,,
\end{equation}
which represents equilateral triangle configurations,
and its ratio to the power spectrum 
\begin{equation}
Q_{\rm eq}(k) \equiv {1 \over 3}
\left[ {\Delta_{\rm eq}^2(k) \over \Delta^2(k)} \right]^2\,.
\end{equation}
In second order perturbation theory,
\begin{equation}
Q_{\rm eq}^{\rm PT} = 1 - \frac{3}{7}\Omega_m^{-2/63}
\end{equation}
and under hyper-extended perturbation theory (HEPT; \cite{ScoFri99} 1999), 
\begin{equation}
Q_{\rm eq}^{\rm HEPT}(k) = \frac{4 - 2^{n(k)}}{1+2^{n(k)+1}} \, ,
\label{Q3}
\end{equation}
which is claimed to be valid in the deeply nonlinear regime.
Here, $n(k)$ is the {\it linear} power spectral index at $k$. 

\begin{figure}[t]
\centerline{\psfig{file=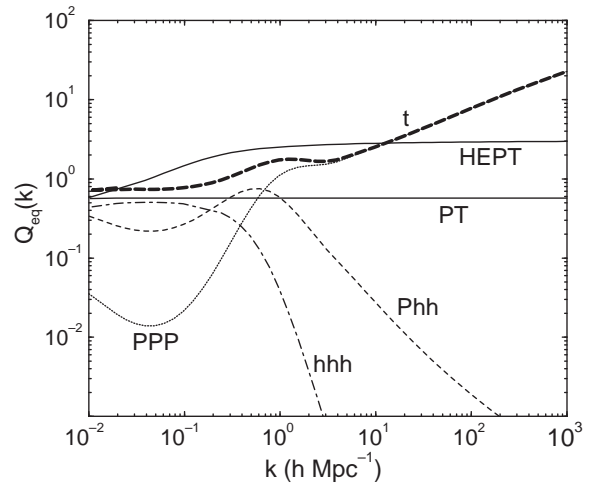,width=3in}}
\caption{$Q_{\rm eq}(k)$ at present broken into individual contributions
under the halo description and compared with second order
perturbation theory (PT) and hyper-extended perturbation theory (HEPT). 
}
\label{fig:dmq}
\end{figure}

In Fig.~\ref{fig:dmpower}(c-d), we show $\Delta_{\rm eq}^2(k)$ 
separated into its various contributions (c) and as a function of
maximum mass (d).  Since the power spectra and equilateral bispectra
share similar features, it is more instructive to examine
$Q_{\rm eq}(k)$ (see Fig.~\ref{fig:dmq}).
Here we also compare it with the second order perturbation theory (PT) and the HEPT prediction.
In the halo prescription, $Q_{\rm eq}$ 
at $k \simgt 10 k_{\rm nonlin} \sim 10 h$Mpc$^{-1}$ 
arises mainly from the single halo term. The HEPT prediction exceeds
the halo prediction on larger scales and falls short on smaller scales.

\subsection{Discussion}

Even though the dark matter halo formalism provides a physically motivated
means for calculating the statistics of the dark matter density field, 
there are several limitations of the approach that should be borne in
mind when interpreting the results.  

The approach assumes all halos to be spherical with a single
profile shape. Any variations in the profile through halo mergers and resulting
substructure can affect the power spectrum and higher order
correlations. Also, real halos are not perfectly
spherical which affects the configuration dependence of the bispectrum.  

Furthermore, there are parameter degeneracies in the formalism that prevent
a straightforward interpretation of observations in terms of halo properties.
For example, one might think that the power spectrum and bispectrum 
can be used to measure any mean deviation from the assumed NFW profile form.
However as pointed out by
\cite{Sel00} (2000), changes in the slope of the inner profile 
can be compensated by changing the concentration as a function of
mass; this degeneracy is also preserved in the bispectrum. 

We do not expect these issues to affect our qualitative results.  However,
if this technique is to be used for precision studies of cosmological 
parameters,
more work will be required in testing it quantitatively against simulations.
Studies by \cite{MaFry00a} (2000a) and \cite{Scoetal00} (2000) show that
the bispectrum predictions of the halo formalism are in good agreement  
with simulations, at least when averaged over configurations.
The replacement of individual halos found in numerical simulations
with synthetic smooth halos with NFW profiles by \cite{MaFry00b}
(2000b) show that the smooth profiles can regenerate the measured
power spectrum and bispectrum in simulations. This agreement, at least
at scales less than 10$k_{\rm nonlin}$, 
suggests that mergers and substructures may not be important at such scales. 

\begin{figure*}[t]
\centerline{\psfig{file=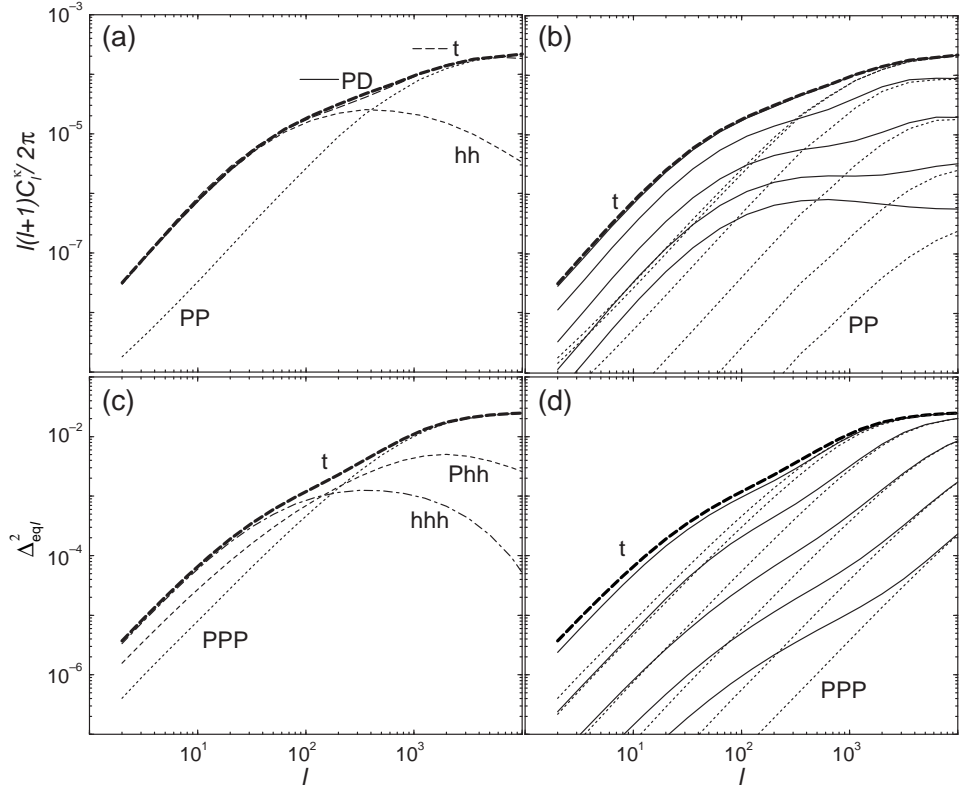,width=5in}}
\caption{Weak lensing convergence spectra under the halo
description.
(a) Angular power spectrum broken into individual contributions.
(b) Mass cut off effects on the angular power spectrum. 
(c) Equilateral bispectrum broken into individual contributions.
(d) Mass cut off effects on the equilateral bispectrum.
Also shown in (a) is the prediction from the PD nonlinear power
spectrum fitting function.  
The mass cuts are the same as in Fig.~\protect\ref{fig:dmpower}
and we have assumed that all sources are at $z_s=1$.
}
\label{fig:lenspower}
\end{figure*}

\section{Convergence Power Spectrum and Bispectrum}
\label{sec:convergence}

\subsection{Power Spectrum and Variance}

The angular power spectrum of the convergence is defined in
terms of the multipole moments $\kappa_{l m}$ as
\begin{equation}
\left< \kappa_{l m}^* \kappa_{l' m'}\right> = C_l^\kappa \delta_{l l'} \delta_{m m'}\,.
\end{equation}
$C_l$ is numerically equal to the flat-sky power spectrum
in the flat sky limit.
It is related to the dark matter power spectrum by (\cite{Kai92} 1992; 1998)
\begin{equation}
C^\kappa_l = \int d\rad \frac{W(\rad)^2}{d_A^2} 
P^\tot\left(\frac{l}{d_A};\rad\right) \, ,
\label{eqn:lenspower}
\end{equation}
where $\rad$ is the comoving distance and  $d_A$ is the angular diameter
distance.  When all background sources are at a distance of 
$\rad_s$, the weight function becomes
\begin{equation}
W(\rad) = \frac{3}{2} \Omega_m \frac{H_0^2}{c^2 a} \frac{
d_A(\rad) d_A(\rad_s -\rad)}{d_A(\rad_s)} \, ; 
\end{equation}
for simplicity, we will assume $\rad_s = r(z_s=1)$ throughout.
In deriving Eq.~(\ref{eqn:lenspower}), we have used the
Limber approximation (\cite{Lim54} 1954) by setting $k=l/d_A$ and
the flat-sky approximation.
In \cite{Cooetal00b} (2000b), we used the projected mass of individual
halos to construct the weak lensing power spectrum directly.
The two approaches are essentially the same since the order in
which the projection is taken does not matter. 

In Fig.~\ref{fig:lenspower}, we show the convergence power
spectrum of the dark matter halos compared with that predicted by
the \cite{PeaDod96} (1996) power spectrum.
The lensing power spectrum due to halos has the same behavior as the
dark matter power spectrum. At large angles ($l \lesssim 100$),
the correlations between halos dominate.
The transition from linear to non-linear is at $l \sim
500$ where halos of mass similar to $M_{\star}(z)$ contribute.
The single halo contributions start dominating at $l > 1000$.

As shown in
Fig.~\ref{fig:lenspower}(b), and discussed in \cite{Cooetal00b}
(2000b), if there is a lack of massive halos in the observed fields
convergence measurements will be biased low compared with the cosmic 
mean.  The lack of massive halos affect the
single halo contribution more than the halo-halo correlation term, thereby
changing the shape of the total power spectrum in addition to
decreasing the overall amplitude.

Similar statements apply to variance statistics (second moments) 
in real space.  The variance of a map smoothed with a window is
related to the power spectrum by
\begin{equation}
\left< \kappa^2(\sigma) \right> =
{1 \over 4\pi} \sum_l (2l+1) C_l^\kappa W_l^2(\sigma)\,.
\label{eqn:secondmom}
\end{equation}
where $W_l$ are the multipole moments (or Fourier transform in a
flat-sky approximation) of the window.   For simplicity, we will choose 
a window which is a two-dimensional top hat in real space with a 
window function in
multipole space of $W_l(\sigma) = 2J_1(x)/x$ with $x = l\sigma$.

\begin{figure*}[t]
\centerline{\psfig{file=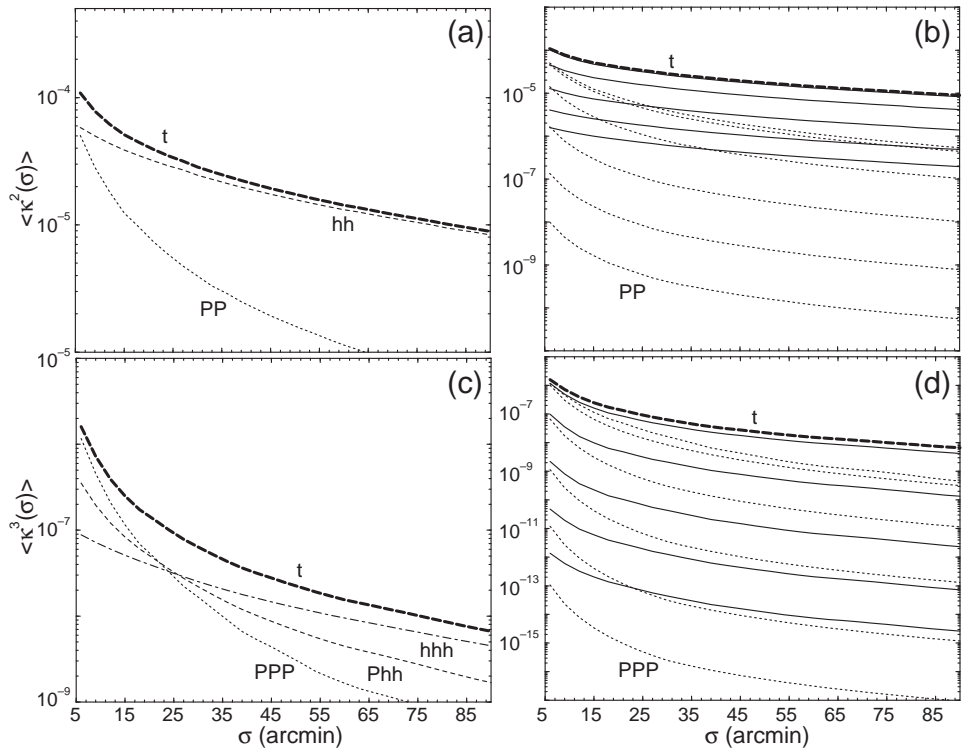,width=5in}}
\caption{Moments of the convergence field as a function of
top-hat smoothing scale $\sigma$. 
(a) Second moment broken into individual contributions. 
(b) Mass cut off effects on the second moments.
(a) Third moment broken into individual contributions. 
(b) Mass cut off effects on the third moments.
The mass cuts are the same as in Fig.~\protect\ref{fig:dmpower}.
}
\label{fig:moments}
\end{figure*}

In Fig.~\ref{fig:moments}(a-b), we show the second moment as a function of
smoothing scale $\sigma$. Here, we have considered angular scales
ranging from 5$'$ to 90$'$, which are likely to be probed by ongoing
and upcoming weak lensing experiments.
As shown, most of the contribution to the second
moment comes from the double halo correlation term and is mildly affected by a 
mass cut off. 
  
\subsection{Bispectrum and Skewness}

The angular bispectrum of the convergence is defined as
\begin{equation}
\left< \kappa_{l_1 m_1} \kappa_{l_2 m_2} \kappa_{l_3 m_3} \right> 
= \wjm B_{l_1 l_2 l_3}^\kappa \,.
\end{equation}
Extending our derivation of the SZ bispectrum in
\cite{Cooetal00b} (2000b), we can write the angular bispectrum of the
convergence as
\begin{eqnarray}
\bi^\kappa &=& \sqrt{(2l_1+1)(2l_2+1)(2l_3+1)  \over 4\pi} \wj
        \nonumber\\
&&\times \left[ \int dr {[W(r)]^3 \over \da^4}  B^\tot\left({l_1 \over
        \da},{l_2 \over \da},{l_3\over \da};r\right)\right] \, .
\label{eqn:szbispectrum}
\end{eqnarray}
The more familiar flat-sky bispectrum is simply the expression in brackets
(\cite{Hu00} 2000). 
The basic properties of Wigner-3$j$ symbol
introduced above can be found in \cite{Cooetal00a} (2000a).

Similar to the density field bispectrum,
we define
\begin{equation}
\Delta^2_{{\rm eq}l} = \frac{l^2}{2 \pi}
\sqrt{B^\kappa_{l l l}} \, ,
\end{equation}
involving equilateral triangles in $l$-space.

In Fig.~\ref{fig:lenspower}(c-d), we show $\Delta^2_{{\rm eq}l}$. 
The general behavior of the lensing bispectrum can be
understood through the individual contributions to the
density field bispectrum: at small multipoles, the triple halo 
correlation term  dominates, while at high multipoles,
the single halo term dominates. The double halo term
contributes at intermediate $l$'s corresponding to angular scales of a
few tens of arcminutes. 

\begin{figure}[b]
\centerline{\psfig{file=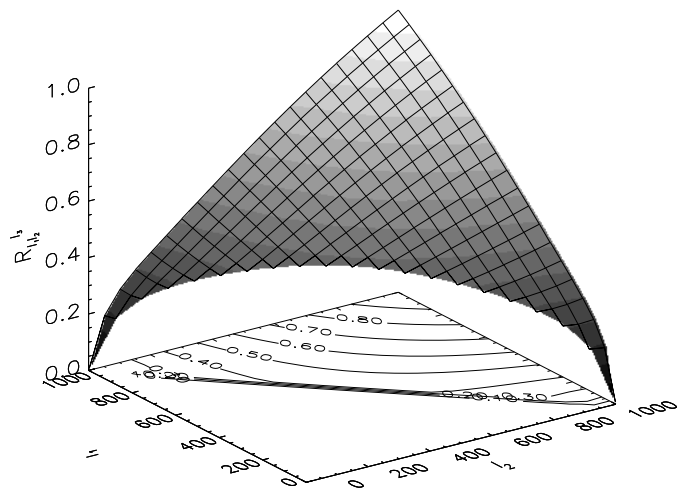,width=4.0in,angle=90}}
\caption{The bispectrum configuration dependence $R_{l_1l_2}^{l_3}$
 as a function of $l_1$ and $l_2$ with
$l_3=1000$.  Due to triangular conditions associated with $l$'s, only the 
upper triangular region in $l_1$-$l_2$ space contribute to the
bispectrum.}
\label{fig:bispecsurface}
\end{figure}

In Fig.~\ref{fig:bispecsurface}, we show the configuration dependence
\begin{equation}
R_{l_1 l_2}^{l_3} = {l_1 l_2\over 2\pi} 
	{\sqrt{B^\kappa_{l_1 l_2 l_3}} \over \Delta_{{\rm eq}l}^2}  \,
\end{equation}
as a function of $l_1$ and $l_2$ when $l_3=1000$.
The surface, and associated contour plot, shows the contribution 
to bispectrum from triangular configurations in $l$ space
relative to that from the equilateral configuration. 
Due to the triangular conditions associated with $l$'s, only
the upper triangular region of $l_1$-$l_2$ space contribute to the
bispectrum. The symmetry about $l_1=l_2$ line is due to the intrinsic
symmetry associated with the bispectrum.
Though the weak lensing bispectrum peaks for equilateral configurations,
the configuration dependence is weak.

The skewness is simply one, easily measured, aspect of the bispectrum.  
It is associated with the third moment 
of the smoothed map (c.f. eqn.~[\ref{eqn:secondmom}])
\begin{eqnarray}
\left< \kappa^3(\sigma) \right> &=&
                {1 \over 4\pi} \sum_{l_1 l_2 l_3}
                \sqrt{(2l_1+1)(2l_2+1)(2l_3+1) \over 4\pi} \nonumber\\
                &&\times \wj \bi^\kappa
                W_{l_1}(\sigma)W_{l_2}(\sigma)W_{l_3}(\sigma)
                \,. \nonumber \\
\end{eqnarray}
We then construct the skewness as
\begin{equation}
S_3(\sigma) =
\frac{\left<\kappa^3(\sigma)\right>}{\left<\kappa^2(\sigma)\right>^2}
\, .
\end{equation}

The effect of the mass cut off is dramatic in the third moment.
As shown  in Fig~\ref{fig:moments}(c-d), most of the contributions to the
third moment come from the single halo term, with those 
involving halo  correlations contributing
significantly only at angular scales greater than $\sim$ 25$'$.
With a mass cut off, 
the total third moment decreases rapidly and is suppressed
by more than three orders of magnitude when the maximum mass drops to 
$10^{13}$ M$_{\sun}$. The skewness only saturates
when the maximum mass is raised to a few times $10^{15}$
M$_{\sun}$. Even though a small change in the maximum mass does not
greatly change the convergence power spectrum (Fig.~3 of
\cite{Cooetal00b} 2000b), the third moment, or the bispectrum, is
strongly sensitive to the rarest or most massive dark matter halos.

\begin{figure}[bt]
\centerline{\psfig{file=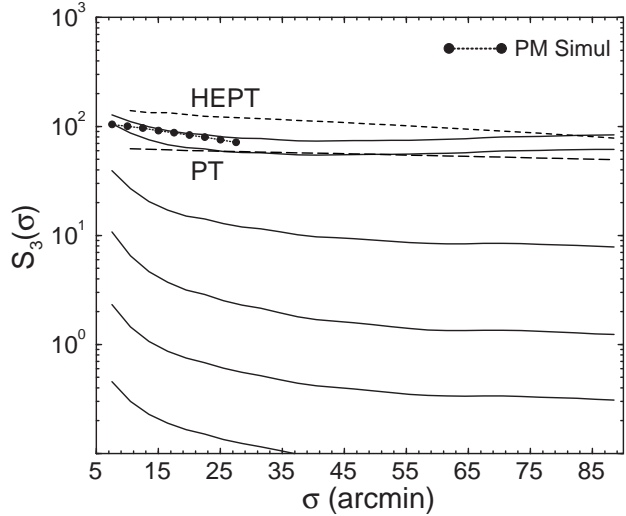,width=3.25in}}
\caption{The skewness, $S_3(\sigma)$, as a function of angular scale.
Shown here is the skewness values with varying maximum mass 
as in Fig.~\protect\ref{fig:dmpower}.
For comparison, we also show
skewness values as measured in particle-mesh (PM) simulations of
\cite{WhiHu99} (1999), as predicted by hyper-extended perturbation
theory (HEPT; dashed line) and
second-order perturbation theory (PT; long-dashed line).} 
\label{fig:skewness}
\end{figure}

In Fig.~\ref{fig:skewness} we plot the  
skewness as a function of maximum mass, ranging from
$10^{11}$ to $10^{16}$ M$_{\sun}$. 
Our total maximum skewness agrees with what is predicted by numerical
particle mesh simulations (\cite{WhiHu99} 1999) and yields 
a value of $\sim$ 116 at 10$'$. 
However, it is lower than predicted by HEPT arguments and 
simulations of \cite{JaiSelWhi00} 
(2000), which suggest a skewness of $\sim$ 140 at angular 
scales of 10$'$. The skewness based on second-order PT is factor of
$\sim$ 2 lower than the maximum skewness predicted by halo
calculation. As shown, the PT skewness decreases slightly from angular
scales of few arcmins to 90$'$ and increases thereafter. 

The effect of maximum mass on the skewness is interesting. 
When the maximum mass is decreased to
$10^{15}$ M$_{\sun}$ from the maximum mass value where skewness
saturates ($\sim 3\times10^{15}$ M$_{\sun}$), 
the skewness decreases from $\sim$ 116 to 98 at an angular scale of
10$'$, though the convergence power spectrum only changes by less than
few percent when the same change on the maximum mass used 
is made. 
When the maximum mass used in the calculation is
$10^{13}$ M$_{\sun}$, the skewness at 10$'$ is $\sim 8$, which is
roughly a factor of 15 decrease in the skewness from the total.

The variation in skewness as a function of angular scale is due to
the individual contribution to the second and third moments. The increase
in the skewness at angular scales less than $\sim$ 30$'$ is
due to the single halo contributions for the third moment.
 The triple halo correlation terms dominate angular scales greater than 50$'$, 
leading to a slight increase toward large angles, e.g.  from
$\sim$ 74 at 40$'$ to $\sim$ 85 at 90$'$. However, this increase is
not present when the 
maximum mass used in the calculation is less than $\sim 10^{14}$ M$_{\sun}$.
Even though mass cut off affects the single halo contributions more than
the halo contribution, at such masses, the change in halo contribution
with mass cut off prevents an increase in skewness at large angular scales.

The absence of rare and massive halos in observed fields will certainly bias 
the skewness measurement from the cosmological mean.  One therefore needs
to exercise caution in using the skewness to constrain cosmological
models \cite{Hui99} 1999). 
In \cite{Cooetal00b} (2000b), we suggested that lensing observations
in a field of $\sim$ 30 deg$^2$ may be adequate for an unbiased
measurement of the convergence power spectrum. For the skewness,
observations within a similar area may be biased by as much as
$\sim$ 25\%.  This is consistent with the sampling errors found in
numerical simulations: 1$\sigma$ errors of 24\% at $10'$ with a 36
deg$^{2}$ field (\cite{WhiHu99} 1999).
To obtain the skewness within few percent of the total,
one requires a fair sample of halos out to
$\sim 3 \times 10^{15}$ M$_{\sun}$,  requiring observations
of $\sim$ 1000 deg$^2$, which is within the reach of upcoming lensing
surveys involving wide-field cameras,
such as the MEGACAM at Canada-France-Hawaii-Telescope
(\cite{Bouetal98} 1998),
 and proposed dedicated telescopes (e.g., Dark Matter Telescope;
Tyson, private communication).

Still, this does not mean that non-Gaussianity measured in smaller fields
will be useless.   With this halo approach one can calculate the expected
skewness if one knows that the most massive halos are not present in 
the observed fields. 
This knowledge may come from external information such as X-ray data and
Sunyaev-Zel'dovich measurements or internally from the lensing data. 

\begin{figure}[b]
\centerline{\psfig{file=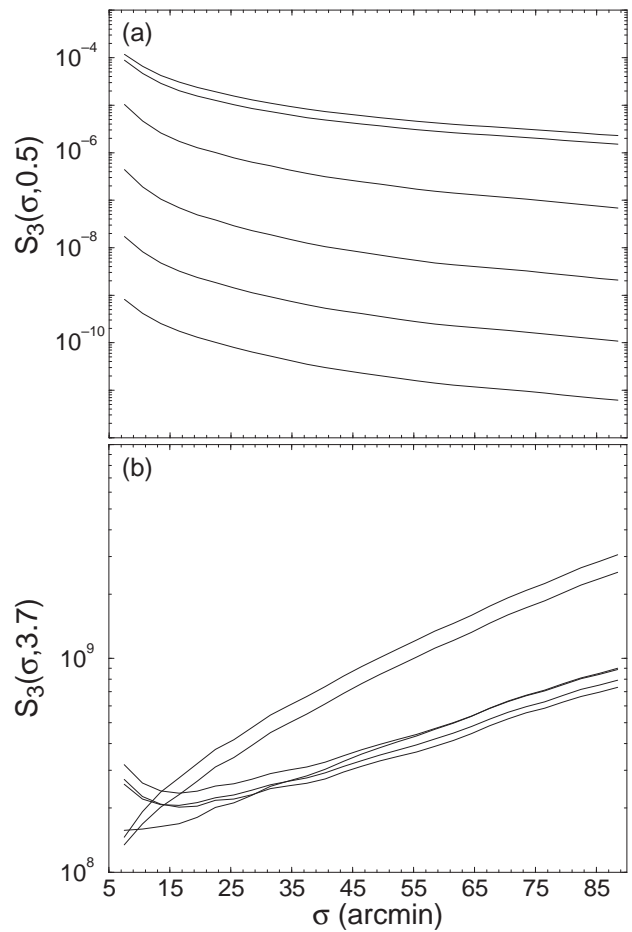,width=3.25in}}
\caption{Generalized skewness statistic $S_3(\sigma,m)$. 
(a) $m=1/2$ following \protect\cite{JaiSelWhi00} (2000).
(b) $m=3.7$ chosen to minimize the mass cut off dependence. 
}
\label{fig:skewmod}
\end{figure}

\subsection{Related Statistics}

The halo description in general allows one to test the effect of rare massive halos 
on any statistic related to the two and three point functions.
In particular, it can be used to design more robust statistics.

Generalized three point statistics have been considered previously by
\cite{JaiSelWhi00} (2000) following \cite{NusDek93} (1993) and 
\cite{Jusetal95} (1995). One such statistic is the $\left< \kappa
|\kappa| \right>_{\kappa>0}$, which is expected to reduce the
sampling variance from rare and massive halos (see, \cite{JaiSelWhi00}
2000 for details). This statistic is proportional to $\left<
\kappa^3\right>/\left< \kappa^2\right>^{1/2}$.
In Fig.~\ref{fig:skewmod}(a), we show this statistic as a function of
maximum mass used in the calculation.  We still find strong 
variations with changes to the maximum mass. Similar variations were
also present in other statistics considered by \cite{JaiSelWhi00}
(2000).

Consider instead the generalized statistic 
\begin{equation}
S_3(\sigma,m) = \left<\kappa^3\right>/\left< \kappa^2 \right>^{m}\,
\end{equation}
where $m$ is an arbitary index. We varied $m$ 
such that the effect of mass cuts
are minimized on skewness. In Fig.~\ref{fig:skewmod}(b), we show such
an example with $m=3.7$. Here, the values are separated to two groups
involving
with most massive and rarest halos and another with halos of masses
$10^{14}$ M$_{\sun}$ or less. Though the values from the two groups
agree with each other on small angular scales, they depart significantly
above $25'$ reaching a difference of 2.5 at 80$'$.  
Statistics involving such a high index $m$,
weight the single halo contributions highly when the most massive halos are
present, whereas they weight the halo correlation terms more strongly
for $M<10^{14}$ M$_{\sun}$.  
To some extent this may be useful to identify the presence of rare
halos in the observations. 

However the consequence of using these generalized statistics
is that one progressively loses their independence on the details
of the cosmological model, e.g. the shape and amplitude of the underlying
density power spectrum, as one departs from $m=2$, thereby contaminating
the probe of dark matter and dark energy.  
Further work is necessary find the optimal trade off between robustness
and cosmological independence of these and other generalized statistics.

\begin{figure}[b]
\centerline{\psfig{file=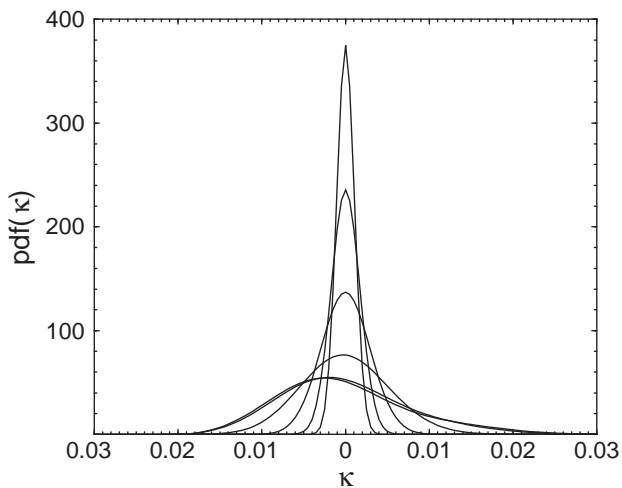,width=3.25in}}
\caption{The probability distribution function of the weak lensing
convergence as a function of maximum mass used in the calculation at
an angular scale of $12'$. From top to bottom, the curves range from
$10^{11}$ to $10^{16}$ M$_{\sun}$.}
\label{fig:pdf}
\end{figure}

Another observable statistic is the probability distribution function 
(pdf) of the convergence maps smoothed on the scale $\sigma$.
This possibility has been recently studied by
\cite{Jaivan99} (1999), where the reconstruction of pdf using peak
statistics were considered. Using the Edgeworth
expansion to capture small deviations from Gaussianity, 
one can write the pdf of convergence to second order as
\begin{eqnarray}
p(\kappa) &=& \frac{1}{\sqrt{2 \pi \left<\kappa^2(\sigma)\right>}} \;
e^{-\kappa(\sigma)^2/2\left<\kappa^2(\sigma)\right>} \\
&& \times \left[1+\frac{1}
{6}S_3(\sigma)\sqrt{\left<
\kappa^2(\sigma)\right>}H_3\left(\frac{\kappa(\sigma)}{\sqrt{\left<\kappa^2(\sigma)\right>}}\right)\right]
\, ,\nonumber
 \end{eqnarray}
where $H_3(x)=x^3-3x$ is the third order Hermite polynomial (see,
\cite{Jusetal95} 1995 for details).  

In Fig.~\ref{fig:pdf}, we show the pdf of convergence at 12$'$ as a
function of maximum mass used in the calculation. As shown, the
greatest departures from Gaussianity begin to occur when the maximum
mass included is greater than $10^{14}$ M$_{\sun}$. Given that we have
only constructed the pdf using terms out to skewness, the presented
pdfs should only be considered as approximate. With increasing
non-Gaussian behavior, the approximated pdfs are likely to depart from
this form especially in the tails.
As studied in \cite{Jaivan99} (1999), the measurement of 
the full pdf can potentially be used a a probe of cosmology.  
Its low order properties describe deviations from Gaussianity near
the peak as opposed to the skewness which is more weighted to the
tails.

\section{Summary \& Conclusions}
\label{sec:conclusions}

We have presented an efficient method to calculate the non-Gaussian
statistics of lensing convergence at the three point level based on
a description of the underlying density field in terms of dark matter halos. 
The bispectrum contains all of the three point information, 
including the skewness.
The prior attempts at calculating lensing bispectrum and
skewness were limited by the accuracy of perturbative approximations and the
dynamic range and sample variance of simulations.  

Though the present technique 
provides a clear and an efficient method
to calculate the statistics of the convergence field,
it has its own shortcomings.  Halos
are not all spherical, which can to some extent affect the
configuration dependence in moments higher than the two point level. 
Substructures due to mergers of halos can
also introduce scatter. Though such effects unlikely to
dominate our calculations, further work using numerical simulations
will be necessary to determine to what extent present method can be
used as a precise tool to study the higher order statistics associated with
weak gravitational lensing.

The dark matter halo approach also allows one to
study possible selection effects that may be present in weak lensing
observations due to the presence or absence of rare massive halos in
the small fields that are observed. 
We have shown that the weak lensing skewness is mostly
due to the most massive and rarest dark matter halos in the
universe. The effect of such halos is stronger at the three point
level than the two point level. The absence of massive halos, with
masses
greater than $10^{14}$ M$_{\sun}$, leads
to a strong decrease in skewness, suggesting that a straightforward
use of measured skewness values as a test of cosmological models may
not be appropriate unless prior observations are available on the
distribution of masses in observed lensing fields.

One can correct for such biases using the halo approach, however. 
 To implement 
such a correction in practice, further work will be needed to calibrate
the technique precisely against simulations across a wide range of
cosmologies. Efficient techniques to correct for mass biases both in
the lensing power spectrum and bispectrum will be needed.
Alternatively, this technique can be used to search for
generalized three point statistics that are more robust to sampling issues.
Given the great potential to study the dark matter distribution through
weak lensing, this issues merit further study.

\acknowledgments
We acknowledge useful discussions with  Lam Hui, Roman Scoccimaro, 
Uros Seljak and Ravi Sheth.
We thank Jordi Miralda-Escud\'e for initial collaborative work that
led to this paper.
ARC acknowledges financial support from John Carsltrom and Don York.
WH is supported by the Keck Foundation.

\end{document}